\documentclass[twoside]{elsart}

\usepackage{amsmath,epsfig,amsfonts,amssymb,latexsym}

\begin{document}
\date{6 December 1999}
\begin{frontmatter}
\title			{Fractional calculus and continuous-time finance}
 
\author[Alessandria]	{Enrico Scalas},
\author[Berlin]        {Rudolf Gorenflo},
and
\author[Bologna]		{Francesco Mainardi}

\address[Alessandria]	{Dipartimento di Scienze e Tecnologie Avanzate, 
			Universit\`a del Piemonte Orientale, 
			via Cavour 84, 
			I--15100 Alessandria, Italy \\
                        and INFN Sezione di Torino, via P.Giuria
                        1, I--10125 Torino, Italy}
\address[Berlin] {Erstes Matematisches Institut, Freie Universitaet Berlin, \\
                 Arnimallee 2-6, D--14195 Berlin, Germany}
\address[Bologna] {Dipartimento di Fisica, Universit\`a di Bologna 
                  and INFN Sezione di Bologna, \\ 
                  via Irnerio 46, I--40126 Bologna, Italy}

\begin{abstract}
In this paper we present a rather general phenomenological theory of
tick-by-tick dynamics in financial markets. Many well-known aspects, such as the
L\'evy scaling form, follow as particular cases of the 
theory. The theory fully takes into account the non-Markovian and
non-local
character of financial time series.
Predictions on the long-time behaviour of the waiting-time 
probability density are presented. 
Finally, a general scaling form is given, based on the
solution of the fractional diffusion equation.
\end{abstract}

\begin{keyword}
Stochastic processes; random walk; statistical finance; econophysics
\\ {{\it PACS: \ }} 02.50.Ey, 02.50.Wp, 89.90.+n
\\ {\it Corresponding author}: Enrico Scalas ({\tt scalas@unipmn.it}), 
\\ url: {\tt www.econophysics.org}
\end{keyword}

\end{frontmatter}

\def\eg{{\it e.g.}\ } \def\ie{{\it i.e.}\ }
\def\sg{\hbox{sign}\,}
\def\sgn{\hbox{sign}\,}
\def\sign{\hbox{sign}\,}
\def\e{\hbox{e}}
\def\exp{\hbox{exp}}
\def\ds{\displaystyle}
\def\dis{\displaystyle}
\def\q{\quad}    \def\qq{\qquad}
\def\lan{\langle}\def\ran{\rangle}
\def\l{\left} \def\r{\right}
\def\lra{\Longleftrightarrow}
\def\arg{\hbox{\rm arg}}
\def\d{\partial}
 \def\dr{\partial r}  \def\dt{\partial t}
\def\dx{\partial x}   \def\dy{\partial y}  \def\dz{\partial z}
\def\rec#1{{1\over{#1}}}
\def\log{\hbox{\rm log}\,}
\def\erf{\hbox{\rm erf}\,}     \def\erfc{\hbox{\rm erfc}\,}
\def\F{\hbox{F}\,}
\def\NN{\hbox{\bf N}}
\def\RR{\hbox{\bf R}}
\def\CC{\hbox{\bf C}}
\def\ZZ{\hbox{\bf Z}}
\def\II{\hbox{\bf I}}


\section{Introduction}

The importance of random walks in finance has been known since the seminal
work of Bachelier \cite{Bachelier 00} which was completed at the end of the 
XIXth century, nearly a hundred years ago. The ideas of Bachelier were further 
carried out by Mandelbrot \cite{Mandelbrot 63}, who introduced the 
concept of L\'evy flights and stable
distributions \cite{Levy 37} in finance, and by the MIT school of 
Samuelson \cite{Merton 90}. 

Although it was well-known that the distribution of returns or of logarithmic returns
approximately followed a stable law, there was a barrier to the application
of these concepts in the financial practice. Indeed, stable distributions have non-finite
variance, and this leads to many mathematical difficulties (for a
discussion on this point the reader is referred to chapter 3 of
Merton's book \cite{Merton 90}). Therefore, in mainstream finance, both
theoreticians and practitioners prefer to use the more tractable continuous 
Wiener process instead of discontinuous L\'evy flights. 
A way of overcoming these difficulties has been provided by empirical studies
suggesting the use of {\em truncated} L\'evy flights, characterized by
probability density distributions with finite moments 
\cite{Mantegna 94,Mantegna 95,Koponen 95,Boyarchenko 99}.

In financial markets, not only prices and returns can be considered
as random variables, but also the waiting time between two transactions varies
randomly. So far, a large part of the financial practice is based on daily
price changes. However, a company specialized in intra--day
transactions and high--frequency data analysis, Olsen \& Associates, has 
published various working papers related to the time behaviour of tick-by-tick data
(see, for instance, on fractional time  ref. \cite{Mueller 93}
and on mean first passage time ref. \cite{Zumbach 98}).

The purpose of this paper is to present a rather general phenomenological theory of
tick-by-tick dynamics in financial markets. Many well-known aspects, such as the
L\'evy scaling form of ref. \cite{Mantegna 95}, follow as particular cases of the 
theory. The theory fully takes into account the non-Markovian and
non-local
character of financial time series.
Predictions on the long-time behaviour of the waiting-time 
probability density are presented. 
Finally, a more general scaling form is given, based on the
solution of the fractional diffusion equation.  

The pa\-per is divided as fol\-lows. In Sec. 2, we 
dis\-cuss the rele\-vance of continuous-\-time
random walks in finance by explicitly performing a mapping from financial
data to random walks. In Sec. 3, we present the master equation and we show that
it reduces to the fractional diffusion equation in the hydrodynamic limit (corresponding
to a long jump-observation scale and long observation times) if some simple
scaling assumptions on 
the jump and waiting-time probability densities hold true. 
Sec. 4 is devoted to the solutions of the fractional diffusion equation and their
natural scaling properties. Finally, in Sec. 5, we point out the main conclusions 
and outline the direction for future work.

As a final remark, let us stress that the theory of continuous-time random walks
is well developed \cite{Montroll 65,Montroll 84}, and its relation to the fractional
diffusion equation and fractional calculus \cite{Samko 93}
has been recently discussed by various authors 
\cite{Fogedby 94,Hilfer 95,Compte 96,Saichev 97,Balescu 97}. 
However, as far as we know, these concepts have not yet been applied to finance in the 
form we present here.

\section{Continuous-time random walk in finance}

The price dynamics in financial markets can be mapped onto a 
random walk whose properties are studied in continuous, rather than
discrete, time \cite{Merton 90}. Here, we shall perform this
mapping, pioneered by Bachelier \cite{Bachelier 00} and fully exploited by
Samuelson and his school \cite{Merton 90}, in a rather general way.

As a matter of fact, there are various ways in which to embed a
random walk in continuous time. Here, we shall base our approach on the
so-called continuous-time random walk (henceforth abbreviated as
{\em CTRW}) in which time intervals between successive steps are 
random variables, as discussed by Montroll and Weiss \cite{Montroll 65}

Let $S(t)$ denote the price of an asset or the value of an index at time
$t$.
In a real market, prices are fixed when demand and offer meet and a 
transaction occurs. In this case,
we say that a trade takes place. In finance, returns rather than 
prices are considered. For this reason, in the following we shall take
into account 
the variable $x(t) = \log S(t)$, that is the logarithm of the price.
Indeed, for a small price variation $\Delta S = S(t_{i+1}) - S(t_{i})$, the 
return $r = \Delta S/S(t_{i})$ and the logarithmic return 
$r_{log} = log[S(t_{i+1}) / S(t_{i})]$ virtually coincide. 

As we mentioned before, 
in financial markets, not only prices can be modelled as random variables, 
but also waiting times between two consecutive 
transactions vary in a stochastic fashion. 
Therefore, the time series $\{ x(t_i) \}$
is characterised by $\varphi(\xi, \tau)$, the 
{\em joint probability density}
of jumps $\xi_{i} = x(t_{i+1}) - x(t_{i})$ and of waiting times 
$\tau_i = t_{i+1} - t_{i}$. The joint density
satisfies the normalization condition 
$\int \int d \xi d \tau \varphi (\xi, \tau) = 1$.

Montroll and Weiss \cite{Montroll 65} have shown that the Fourier-Laplace
transform
of $p(x,t)$, the probability density function, {\em pdf}, of
finding 
the value $x$ of the price logarithm (which is the diffusing quantity in 
our case) at time $t$, is given by:
\begin{equation}
\label{montroll}
\widetilde{\widehat p}(\kappa, s) = {1-\widetilde \psi(s) \over s}\,
{1 \over 1- \widetilde{\widehat \varphi}(\kappa, s)}\,,
\end{equation}
where 
\begin{equation}
\label{transform}
\widetilde{\widehat p}(\kappa, s) = 
\int_{0}^{+ \infty} dt \; \int_{- \infty}^{+ \infty} dx \, 
\e^{\,\ds -st+i \kappa x} \, p(x,t)\,,
\end{equation}
and $\psi(\tau) = \int d \xi \, \varphi(\xi, \tau)$ 
is the waiting time pdf.

Let us now consider the situation in which the waiting time and the size
of the step are {\em independent}. In this case the joint 
density function, $\varphi$, can be factorized, namely written as the
product of a 
``spatial'' part and a temporal part: 
$\varphi(\xi, \tau) = \lambda(\xi) \psi (\tau)$. 
Here $\lambda (\xi)$  is the probability for a displacement $\xi$
in each single step (transition probability density). Now, the normalization
condition for the transition pdf: $\int d \xi \lambda (\xi) 
= 1$ must be added to that for the probability density of the 
waiting time $\int d \tau \psi(\tau)=1$. 

As a consequence we get:
\begin{equation}
\label{montroll2}
\widetilde{\widehat p}(\kappa ,s) = {1-\widetilde\psi(s)  \over s}\,
 {1 \over 1- \widehat \lambda (\kappa)\, \widetilde \psi(s)}
 = {\widetilde {\Psi}(s)  \over
   1 - \widehat \lambda (\kappa)\, \widetilde \psi(s)}
\,,
\end{equation}
where $\widehat \lambda (\kappa)\,,$ the Fourier transform of
the transition probability density, is usually called
the {\it structure function} of the random walk and
 $\widetilde {\Psi}(s) =(1 - \tilde{\psi}(s) ) / s$ is the Laplace
transform of
\begin{equation}
\label{survival} 
\Psi(t) = \int_t^{\infty} \psi(t')\, dt' =
1- \int_0^t \psi(t')\, dt'\,.
\end{equation}
$\Psi(t)$ is the {\em survival probability}  at the initial point position
($t_0 = 0$) \cite{Hilfer 95}.
$\int_0^t \psi(t')\, dt'\,$ represents the probability that at least one
step is taken in the interval ($0,t$), hence
$\Psi(t)\,$   is the probability that the diffusing quantity
does not change during the time interval 
of duration $t$ after a jump \cite{Balescu 97}.

According to Weiss \cite{Weiss 94}, $\Psi(t)$ can be viewed as the probability
that the duration of a given interval between successive steps is strictly
greater than 
$t$ and is the peculiar function needed to specify the probability of the
displacement at time $t^{*}+t$ in a CTRW, where $t^*$ is the instant
of the last jump. The waiting-time  pdf
is related to $\Psi(t)$ by the  formula: $\psi (t) = -
d \Psi(t)/ dt$.

Let us finally remark that, in
general, the CTRW is a non-Markovian model \cite{Weiss 94}, as at any time one has to know
the value of the diffusing quantity as well as the time at which the last
step took place in order to predict the further course of the walk.
The non-Markovian property arises because the time of the previous step
does vary and could be even $t=0$, so that the complete history of the
process must be taken into account at all times. The only Markovian version of the
CTRW is the one in which the waiting time pdf, $\psi(\tau)$, is a negative exponential:
$$
\psi(\tau) = \frac{1}{T} \exp (-\tau/T),
$$
where $T$ is the average time between successive steps.
Only for this form of the density, the probability that
a step of the random walk will take place
in $(t\,,\,t +dt)$ is $dt/T\,, $ as $dt \to 0\,, $
independent of the time at which the immediately preceding
step occurred. 
This is not true of any other form of $\psi(\tau)\,. $

\section{Master equations and fractional diffusion}

The {\em master equation} governing the probability
density profile in a CTRW can be derived by inverting the Fourier-Laplace
transform in eq. (\ref{montroll2}). Rewriting (\ref{montroll2}) as
$$
\widetilde{\widehat p}(\kappa ,s) = \widetilde{\Psi}(s) + 
\widetilde{\psi}(s) \widehat \lambda (\kappa) \widetilde{\widehat p}(\kappa ,s)
$$
we obtain
\begin{equation}
\label{master1}
p(x,t) =  \delta _{x\,0} \Psi(t) +
   \int_0^t dt' \,  \psi(t-t') \,
      \int_{-\infty}^{+\infty}  dx'\, \lambda (x-x')\,
    p(x',t')\,.
\end{equation}
This form of the master equation is quoted, e.g., in Klafter et al. 
\cite{Klafter 87} and Hilfer and Anton \cite{Hilfer 95}. However, equivalent
forms can be found in the literature. The following form shows the non-local
and non-Markovian character of the CTRW \cite{Montroll 84,Balescu 97}:
\begin{equation}
\label{master2}
{\d \over \d t} p(x,t) =
   \int_0^t dt' \,  \phi(t-t') \, \l [
      -p(x,t') + \int_{-\infty}^{+\infty}  dx'\, \lambda (x-x')\,
  p(x',t')\r]\,;
\end{equation}
here, the kernel $\phi(t)$ is defined through its Laplace transform
$$ 
\widetilde{\phi} (s) =
{ s\, \widetilde{\psi}(s) \over  1- \widetilde{\psi}(s)}\,.
$$

The above equations allow to compute $p(x,t)$ from the knowledge
of the jump pdf $\lambda(\xi)$ 
and of the waiting-time pdf
$\psi(\tau)$. In principle, both these quantities are empirically accessible 
from high-frequency market data, even if, recently, within the physics 
community, emphasis has been given to the jump pdf \cite{Mantegna 95}.
 
The time-evolution equation for $p(x,t)$ has a remarkable limit, 
if some scaling conditions
on the structure function and on the waiting time pdf are verified.
 
Let us assume the following scaling behaviour in the hydrodynamic 
limit (long-jump scale and long observation times):
\begin{equation}
\label{scalingjumps}
\widehat \lambda(\kappa ) \sim 1 - |\kappa |^{\alpha}\,,
 \q \kappa  \to 0\,, \q 0<\alpha \le 2\,,
\end{equation}
and 
\begin{equation}
\label{scalingtime}
\widetilde \psi(s) \sim 1 - s^{\beta}    \,, \q s \to 0\,,
 \q 0<\beta \le 1\,.
\end{equation}

The above approximations are consistent with the following explicit
expressions for the Fourier and Laplace transforms:
\begin{equation}
\label{scalingjumps2}
\widehat \lambda (\kappa ) = \exp (-|\kappa |^\alpha)\,,
\q 0<\alpha \le 2\,,
\end{equation}
and
\begin{equation}
\label{scalingtime2}
\widetilde \psi(s)= {1 \over 1+  s^{\beta}}
   \,, \q 0<\beta \le 1\,.
\end{equation}
We note that  eq. (\ref{scalingjumps2}) represents the characteristic function for the
symmetric L\'evy stable pdf of index $\alpha \,; $  for $0<\alpha <2$
the pdf decays like $|x|^{-(\alpha +1)}$ as $|x| \to \infty\,, $
for $\alpha =2$
the Gaussian pdf is recovered.

From eq. (\ref{scalingtime2}), we observe that
\begin{equation}
\label{mittag1}
\widetilde\Psi(s) = {1-\widetilde\psi(s)  \over s} =
 {s^{\beta-1} \over 1+ s^\beta}\,,   \q 0<\beta \le 1\,,
\end{equation}
so that the survival probability  turns out to be
\begin{equation}
\label{mittag2}
\Psi(t) = E_\beta (-t^\beta)
\,,   \q 0<\beta \le 1\,,
\end{equation}
where
$$  E_\beta (-t^\beta) =  \sum_{n=0}^{\infty}
  (-1)^n\, { t^{\beta n} \over \Gamma(\beta n + 1)} $$
is the Mittag-Leffler function of order $\beta $
\cite{Erdelyi 55,Gorenflo 97}.
Thus  the pdf for the waiting time is
\begin{equation}
\label{waitingpdf} 
\psi(t) = -   {d \over dt} \Psi(t) =
            -   {d \over dt}  E_\beta (-t^\beta )
 \,, \q 0<\beta \le 1\,,
\end{equation}
which is in agreement with the expression obtained in
\cite{Hilfer 95} in terms of the generalized Mittag-Leffler function
in two parameters.

For $0<\beta <1$ the Mittag-Leffler function
$E_\beta(-t^\beta)$ is known to be, for $t>0$, a
completely monotonic function of $t$, decreasing from 1 (at $t=0$)
to 0 like $t^{-\beta}\,$ as $t \to \infty\,$ \cite{Gorenflo 97}.
As a consequence the pdf for the waiting time is
strictly positive and monotonically decreasing to zero like
$t^{-(\beta +1)}\,.$
For $\beta =1$ the Mittag-Leffler function reduces
to $\exp (-t)\,$ and
we recover from  eqs. (\ref{mittag1} -- \ref{waitingpdf})  the
Markovian CTRW.

If we insert eqs. (\ref{scalingjumps}) and (\ref{scalingtime})
into eq. (\ref{montroll}), we get the limiting relation:
\begin{equation}
\label{pseudodifferential}
s^{\beta} \widetilde {\widehat p}(\kappa,s) + |\kappa|^{\alpha} 
\widetilde {\widehat p}(\kappa,s) = s^{\beta -1}\,.
\end{equation}

Inverting eq. (\ref{pseudodifferential}), we obtain the time-evolution
equation for $p(x,t)$ in the hydrodynamic limit.
If $0 < \beta \leq 1$ and $0 < \alpha \leq 2$, we have, for $x \in R$:
\begin{equation} 
\label{fractional}
\frac{\partial^{\beta} p(x,t)}{\partial t^{\beta}} =
\frac{\partial^{\alpha} p(x,t)}{\partial |x|^{\alpha}} +
\frac{t^{-\beta}}{\Gamma(1-\beta)} \delta(x), \; \; (t>0).
\end{equation}

In eq. (\ref{fractional}), we have introduced the fractional derivatives
$\partial^{\beta} / \partial t^{\beta}$ and $\partial^{\alpha} / 
\partial |x|^{\alpha}$ defined as the 
inverse Laplace and Fourier transforms
of $s^{\beta}$ and $-|\kappa|^{\alpha}$, respectively 
\cite{Samko 93,Saichev 97}.
Fractional derivatives are non-local operators belonging to the 
larger class of pseudo-differential operators 
\cite{Hoermander 85,Jacob 96}, which
allow power-law effects. In particular, the ``time''
operator in eq. (\ref{fractional}) is the Riemann-Liouville
fractional derivative of order $\beta$ defined as 
(if $0< \beta < 1$):
$$ 
{d^\beta \over d t^\beta} f(t)=
{1 \over \Gamma(1-\beta )}\,{d \over d t} \l\{\int_0^t
{f(\tau )\over (t-\tau )^{\beta}} \,d \tau \r\} \,,
$$
whereas the ``jump'' operator is the Riesz fractional derivative of order
$\alpha$ which, if $0 < \alpha < 2$ can be represented as \cite{Samko 93}:
$$ {d^\alpha  \over d |x|^\alpha} f(x)	=
 \Gamma(1+\alpha ) \,
 {\sin \,(\alpha \pi/2) \over \pi }\,
 \int_0^\infty
 {f(x+\xi)- 2f(x) + f(x-\xi) \over {\xi}^{1+\alpha}}\, d \xi
 \,.$$

Finally, let us mention that eq. (\ref{pseudodifferential})
was derived by Weiss \cite{Weiss 94} and by Afanas'ev and co-workers
\cite{Afanas'ev 91}. Moreover, the above derivation of eq.
(\ref{fractional}) was implicit in a paper by Fogedby \cite{Fogedby 94}
and was explicitly presented by Compte \cite{Compte 96} and by Saichev and
Zaslavsky \cite{Saichev 97}.

\section{L\'evy flights and scaling of solutions}

We start this section with the analysis of a particular case of eq. 
(\ref{fractional}), the limit $\beta \to 1$, where we have (in the weak
sense) \cite{Saichev 97}:
$$
\lim_{\beta \to 1} \frac{t^{-\beta}}{\Gamma (1 - \beta)} = \delta(t),
$$
and eq. (\ref{fractional}) becomes equivalent to the following initial value
problem:
\begin{equation}
\label{cauchylevy}
\frac{\partial p(x,t)}{\partial t} = 
\frac{\partial^{\alpha} p(x,t)}{\partial |x|^{\alpha}},\, \q
p(x,0) = \delta(x).
\end{equation}
The Cauchy problem (\ref{cauchylevy}) can be solved by 
Fourier-trans\-forming both
sides of the equation with respect to $x$. After integrating and inverse
Fourier-trans\-forming, one gets:
\begin{equation}
\label{levypdf}
p(x,t) = \frac{1}{t^{1/\alpha}} L_{\alpha} \left ( \frac{x}{t^{1/\alpha}} 
\right ),
\end{equation}
where $L_{\alpha} (u)$
is the L\'evy standardized probability density function:
\begin{equation}
\label{levyscaling}
L_{\alpha} (u) = \frac{1}{2 \pi} \int_{-\infty}^{+\infty}
\e^{-iqu-|q|^{\alpha}} dq.
\end{equation}
Taking the limit $\beta \to 1$ in eq. (\ref{fractional}) corresponds to 
considering independent time increments. Continuous-time random walks 
whose pdf $p(x,t)$ is given by eq. (\ref{levypdf}) are called 
{\em symmetric L\'evy flights} or better {\em symmetric $\alpha$-stable
L\'evy processes} \cite{Levy 37,Janicki 94}. In 1963, analysing the 
scaling properties of financial time-series, Mandelbrot 
\cite{Mandelbrot 63} found that the empirical pdf $p(x,t)$ could be well fitted 
by the L\'evy density function (\ref{levypdf}) with $\alpha = 1.7$. 
As we mentioned in the introduction,
the main difficulty in dealing with the L\'evy distribution is that its
moments diverge. For $0<\alpha<2$, the only bounded finite moments
have index $\gamma$ satisfying $-1 < \gamma < \alpha$. For this reason, the
results of Mandelbrot were well-known but not much used in mainstream
quantitative finance \cite{Merton 90}. The recent empirical analysis of 
Mantegna and Stanley \cite{Mantegna 95} suggests that {\em truncated} L\'evy 
flights should be used instead, as good models for financial price dynamics 
\cite{Mantegna 94}.
Koponen \cite{Koponen 95} introduced a class of truncated L\'evy flights, which
was successively generalized by Boyarchenko and Levendorskii 
\cite{Boyarchenko 99}. However, all these studies somehow neglected the 
waiting-time pdf. 

In the general case, the Cauchy problem of eq. (\ref{fractional}) can be solved 
by the same technique used above. There is, however, a mathematical subtlety.
In order to give a meaning to the Cauchy problem, the Riemann-Liouville
operator must be replaced by the Caputo fractional derivative of order
$\beta$ \cite{Gorenflo 97,Podlubny 99}:
$$ {d^\beta \over d t^\beta} f(t)=
  {1 \over \Gamma(1-\beta )}\,{d \over d t} \l\{
    \int_0^t
  {f(\tau )\over (t-\tau )^{\beta}} \,d \tau \r\}
  - {t^{-\beta }\over \Gamma(1-\beta)}\, f(0) \,.
$$
Now, the solution is:
\begin{equation}
\label{generalpdf}
p(x,t) = \frac{1}{t^{\beta/\alpha}} W_{\alpha,\beta} \left ( \frac{x}{t^{\beta/\alpha}} 
\right ),
\end{equation}
and $W_{\alpha,\beta} (u)$ is the following scaling function:
\begin{equation}
\label{generalscaling}
W_{\alpha,\beta} (u) = \frac{1}{2 \pi} \int_{-\infty}^{+\infty}
\e^{-iqu} E_{\beta} (-|q|^{\alpha}) dq,
\end{equation}
where $E_{\beta}$ the Mittag-Leffler function of order $\beta $ and argument
$z= - |q|^{\alpha}$

Further empirical studies on high-frequency financial data may reveal
the scaling form (\ref{generalscaling}), if the waiting-time pdf
satisfies the asymptotics (\ref{scalingtime}).

\section{Conclusions and outlook}

In this paper, we argued that the continuous-time random walk (CTRW) is a good 
phenomenological model for high-frequency price dynamics in financial
markets, as, in general, this dynamics is non-Markovian and/or 
non-local.

CTRW naturally leads to the so-called fractional diffusion equation in the hydrodynamic
limit if some scaling properties of the waiting time pdf $\psi(\tau)$ and
of the jump pdf $\lambda (\xi)$ hold true in that limit.
This point needs a 
further discussion. Indeed, the scaling regime of eqs. (\ref{scalingtime}) and
(\ref{scalingjumps}) breaks down for very large jumps. For this reason,
truncated L\'evy flights have been introduced 
\cite{Mantegna 94,Mantegna 95,Koponen 95,Boyarchenko 99}.
Preliminary investigations \cite{Scalas 99} on high frequency financial data show that
a similar problem is present for the waiting time pdf. Nevertheless, 
we can view the fractional diffusion equation (\ref{fractional}) as a model 
for approximating the true behaviour of returns in financial markets.

In the region where the dynamics is well approximated by eq. (\ref{fractional}), we expect 
the following scaling for the waiting-time pdf (see the discussion in Sec. 3):
\begin{equation}
\label{exponent}
\psi(\tau) \sim \tau^{-\mu},
\end{equation}
where $\mu = \beta +1$ varies in the range $1 < \mu < 2$.
Consequently, the more complex scaling form (\ref{generalscaling}) should hold true.

Empirical analyses on market high-frequency data will be necessary in order to verify 
these predictions. In any case, we expect that the concepts of CTRW and of fractional
calculus will be of help in practical applications such as option pricing, as they provide
an intuitive background for dealing with non-Markovian and non-local
random processes.

In this paper, the mathematical apparatus has been kept to a minimum, the interested 
reader will find full mathematical details in a forthcoming paper
\cite{Gorenflo 99}.

\section*{Acknowledgements}
This work was partially supported by the Italian INFM and INFN, and by
the Research Commission of the Free University in Berlin. R.
G. is grateful to the Italian "Istituto Nazionale di Alta Matematica"
for supporting his visit in Italy. E. S. wishes to thank CERN for its wonderful library open
24 hours a day: an ideal place to study in a quiet atmosphere after shifts at the colliders. 
Discussions on high-frequency financial data with Marco Raberto inspired this work.


\begin{thebibliography}{}

\bibitem{Bachelier 00} L.J.B. Bachelier: Theorie de la Speculation, 
Gauthier-Villars, Paris, 1900, Reprinted in 1995, Editions Jaques 
Gabay, Paris, 1995.
\bibitem{Mandelbrot 63}B. Mandelbrot, Journal of Business \textbf{36}
(1963) 394.
\bibitem{Levy 37}P. L\'evy, Th\'eorie de l'Addition des Variables
Al\'eatoires, Gauthier--Villars, Paris, 1937.
\bibitem{Merton 90}R.C. Merton, Continuous Time Finance, Blackwell, 
Cambridge, MA, 1990.
\bibitem{Mantegna 94} R.N. Mantegna, H.E. Stanley, 
Phys. Rev. Lett. {\bf 73} (1994) 2946. 
\bibitem{Mantegna 95} R.N Mantegna and H.E. Stanley, Nature {\bf 376} (1995) 46.
\bibitem{Koponen 95} I. Koponen, Phys. Rev. E {\bf 52} (1995) 1197.
\bibitem{Boyarchenko 99} S.I. Boyarchenko and S.Z. Levendorskii, 1999
preprint, downloadable from:
{\tt http://l3www.cern.ch/\-homepages/susinnog\-/Library/\-biblioindex.html}.
\bibitem{Mueller 93} U.A. M\"uller, M.M. Dacorogna, R.D. Dav\'e, O.V. Pictet, 
R. B. Olsen and J.R. Ward, 1993, Olsen, internal document UAM.1993-08-16, downloadable
from {\tt http://www.olsen.ch/library/research/oa\_working.html}.
\bibitem{Zumbach 98} G. Zumbach, 1998, Olsen, internal document GOZ.1998-01-15, downloadable
from {\tt http://www.olsen.ch/library/research/oa\_working.html}.
\bibitem{Montroll 65} E.W. Montroll and G.H. Weiss, J. Math. Phys. {\bf 6} (1965) 167.
\bibitem{Montroll 84} E.W. Montroll and M.F. Shlesinger,
On the wonderful world of random walks, in: J. Lebowitz and E.W. Montroll
(Eds.), Nonequilibrium Phenomena II: from Stochastics to Hydrodynamics,
North-Holland, Amsterdam, 1984, pp. 1--122.
\bibitem{Samko 93} S.G. Samko, A.A. Kilbas, O.I. Marichev, Fractional
Integrals and Derivatives. Theory and Applications, Gordon and Breach
Science Publishers, 1993.
\bibitem{Fogedby 94} H.C. Fogedby, Phys. Rev. E {\bf 50} (1994) 1657.
\bibitem{Hilfer 95} R. Hilfer and L. Anton, Phys. Rev. E {\bf 51} (1995)
R848.
\bibitem{Compte 96} A. Compte, Phys. Rev. E {\bf 53} (1996) 4191.
\bibitem{Saichev 97} A.I. Saichev and G.M. Zaslavsky, Chaos {\bf 7} (1997) 753.
\bibitem{Balescu 97} R. Balescu, Statistical Dynamics: Matter out of
Equilibrium, Imperial College Press - World Scientific, London, 1994.
\bibitem{Weiss 94} G.H. Weiss, Aspects and Applications of the Random
Walk, North-Holland, Amsterdam, 1994.
\bibitem{Klafter 87} J. Klafter,  A. Blumen and M.F. Shlesinger,
  Phys. Rev. A {\bf 35} (1987)  3081.
\bibitem{Erdelyi 55} A. Erd\'elyi (Ed.), Higher Transcendental Functions,
Bateman Project, Vol. 3, McGraw-Hill, New York, 1955.
\bibitem{Gorenflo 97} R. Gorenflo and F. Mainardi, Fractional calculus,
integral and differential equations of fractional order, in
Fractals and fractional calculus in continuum mechanics, A. Carpinteri
and F. Mainardi (Eds.), Springer, Wien and New York, 1997, pp. 223--276.
\bibitem{Hoermander 85} L. H\"o rmander, The Analysis of Linear
Partial Differential Operators III. Pseudo-Differential Operators,
Springer, Berlin, 1985.
\bibitem{Jacob 96} N. Jacob, Pseudo-Differential Operators and Markov
Processes, Akademie Verlag, Berlin, 1996.
\bibitem{Afanas'ev 91} V.V. Afanas'ev, R.Z. Sagdeev, and G.M. Zaslavsky,
Chaos {\bf 1} (1991) 143.
\bibitem{Janicki 94} A. Janicki, A. Weron, Simulation and Chaotic
Behavior of $\alpha$--Stable Stochastic Processes, Marcel Dekker, New York,
1994.
\bibitem{Podlubny 99} I. Podlubny, Fractional Differential Equations, Academic
Press, San Diego, 1999.
\bibitem{Scalas 99} E. Scalas, M. Raberto, F. Mainardi, and R. Gorenflo, in preparation.
\bibitem{Gorenflo 99} R. Gorenflo, F. Mainardi, and E. Scalas, in preparation.

\end{thebibliography}
\end{document}